\documentclass[amsmath,amssymb,proc,secnumarabic,twocolumn]{revtex4}

\usepackage{graphicx,cancel}
\usepackage{ulem}

\newcommand{\rb}[1]{\raisebox{1.5ex}[-1.5ex]{#1}}

\countdef\preprint=0 \countdef\tube=1
\ifnum\tube<10

\else

\fi

\def\enulld{$E_0+\Delta$}
\def\e0{$E_0$}
\def\eff{$f_1({\bf k})$}
\def\effs{$f_1({\bf k})^*$}
\def\effzwei{$f_2({\bf k})$}
\def\effszwei{$f_2({\bf k})^*$}
\def\effdrei{$f_3({\bf k})$}
\def\effsdrei{$f_3({\bf k})^*$}

\def\sqd{\sqrt{3}}

\begin{document}

\setlength{\mathindent}{0.2cm}

\title{Tight--binding description of the quasiparticle dispersion of graphite and few--layer graphene}

\author{A.~Gr\"{u}neis$^{1,2}$\footnote{Corresponding author.\\Tel.: +49 351 4659 519\\e--mail: ag3@biela.ifw-dresden.de\\(A.~Gr\"{u}neis)$^{1}$},
\emph{}C.~Attaccalite$^{3}$,L.~Wirtz$^{4}$, H.~Shiozawa$^{5}$,
R.~Saito$^6$,T.~Pichler$^{1}$, A.~Rubio$^{3}$}

\affiliation{$^{1}$Faculty of Physics, Vienna University,
Strudlhofgasse 4, 1090 Wien, Austria}

\affiliation{$^{2}$IFW Dresden, P.O. Box 270116, D-01171 Dresden,
Germany}

\affiliation{$^3$Dept. Fisica de Materiales, Donostia International Physics Center, Spain\\
European Theoretical Spectroscopy Facility (ETSF), Spain}

\affiliation{$^{4}$Institute for Electronics, Microelectronics,
and Nanotechnology Dept. ISEN B.P. 60069 59652 Villeneuve d'Ascq
Cedex, France}

\affiliation{$^5$Advanced Technology Institute, University of
Surrey, Guildford, GU2 7XH, UK}

\affiliation{$^6$Department of Physics, Tohoku University, Aoba, Sendai, 980-8578, Japan}

\date{\today}
\begin{abstract}
A universal set of third--nearest neighbour tight--binding~(TB)
parameters is presented for calculation of the quasiparticle~(QP)
dispersion of $N$ stacked $sp^2$ graphene layers ($N=1\ldots
\infty$) with $AB$ stacking sequence. The QP bands are strongly
renormalized by electron--electron interactions which results in a
20\% increase of the nearest neighbour in--plane and
out--of--plane TB parameters when compared to band structure from
density functional theory. With the new set of TB parameters we
determine the Fermi surface and evaluate exciton energies, charge
carrier plasmon frequencies and the conductivities which are
relevant for recent angle--resolved photoemission, optical,
electron energy loss and transport measurements. A comparision of
these quantitities to experiments yields an excellent agreement.
Furthermore we discuss the transition from few layer graphene to
graphite and a semimetal to metal transition in a TB framework.
\end{abstract}
\maketitle

\section{Introduction}
Recently mono-- and few--layer graphene (FLG) in an $AB$ (or
Bernal) stacking is made with high crystallinity by the following
three methods; epitaxial growth on SiC~\cite{seyller06-sic},
chemical vapour deposition on Ni(111)~\cite{alex07-graphenenickel}
and by mechanical cleavage on $\rm SiO_2$~\cite{novoselov05-pnas}.
Graphene is a novel, two--dimensional (2D) and meta stable
material which has sparked interest from both basic science and
application point of view~\cite{geim07-review}. A monolayer of
graphene allows one to treat basic questions of quantum mechanics
such as Dirac Fermions or the Klein
paradox~\cite{geim06-kleinparadox} in a simple condensed--matter
experiment. The existence of a tunable gap in a graphene bilayer
was shown by angle--resolved photoemission
(ARPES)~\cite{rotenberg06-graphite_bilayer}, which offers a
possibility of using these materials as transistors in future
nanoelectronic devices that can be lithographically
patterned~\cite{berger07-graphene}. Furthermore a graphene layer
that is grown epitaxially on a Ni(111) surface is a perfect spin
filter device~\cite{karpan08-spinfilter} that might find
applications in organic spintronics.

It was shown recently by ARPES that the electronic structure of
graphene~\cite{rotenberg06-graphite} and its 3D parent material,
graphite~\cite{alex06-correlation,zhou05-graphite,takahashi07-prl},
is strongly renormalized by correlation effects. To date the best
agreement between ARPES and ab--initio calculations is obtained
for $GW$ (Greens function $G$ of the Coulomb interaction $W$)
calculations of the QP dispersion. The band structure in the local
density approximation (LDA) (bare band dispersion) is not in good
agreement with the ARPES spectra because it does not include
long--range correlation effects. The self energy correction of the
Coulomb interaction to the bare energy band structure are crucial
for determining the transport and optical properties (excitons)
and related condensed--matter phenomena. For graphite, a semi
metal with a tiny Fermi surface, the number of free electrons to
screen the Coulomb interaction is low ($\sim 10^{19}$
carriers~$\rm cm^{-3}$) and thus the electron--electron
correlation is a major contribution to the self--energy
correction~\cite{alex06-correlation}. Theoretically the bare
energy band dispersion is calculated by the local density
approximation (LDA) and the interacting QP dispersion is obtained
by the $GW$ approximation. The $GW$ calculations are
computationally expensive and thus only selected $k$ points have
been calculated~\cite{alex06-correlation}. Therefore a
tight--binding (TB) Hamiltonian with a transferable set of TB
parameters that reproduces the QP dispersion in $sp^2$ stacked
graphene sheets is needed for analysis of ARPES, optical
spectroscopies and transport properties for pristine and doped
graphite and FLGs. So far there are already several sets of TB
parameters published for graphene, FLG and graphite. For graphene
a third nearest neighbour fit to LDA has been
performed~\cite{reich2002}. Recently, however it has been shoen by
ARPES that the LDA underestimates the slope of the bands and also
the trigonal warping effect~\cite{alex06-correlation}. For bilayer
graphene the parameters of the so--called
Slonzcewski--Weiss--McClure (SWMC) Hamiltonian have been fitted to
reproduce double resonance Raman data~\cite{pimenta07-bilayer}. A
direct observation of the quasiparticle (QP) band structure is
possible by ARPES. A set of TB parameters has been fitted to the
experimental ARPES data of graphene grown on SiC~\cite{eli07-ssc}.
As a result they obtained a surprisingly large absolute value of
the nearest neighbour $\pi$ hopping parameter of
5.13~eV~\cite{eli07-ssc}. This is in stark contrast to the fit to
the LDA calculation which gives only about half of this
value~\cite{reich2002}. Considering the wide range of values
reported for the hopping parameters, a reliable and universal set
of TB parameters is needed that can be used to calculate the QP
dispersion of an arbitrary number of graphene layers. The band
structure of FLGs has been calculated~\cite{partons06-graphite}
using first--nearest neighbour in--plane coupling which provide
the correct band structure close to $K$ point. However, as we will
show in detail in this paper, the inclusion of third--nearest
neighbours (3NN) is essential in describing the experimental band
structure in the whole BZ. The fact that the inclusion of 3NN is
essential is also proven by the dispersion of a localized state at
the zig-zag edge of a graphene flake. Only inclusion of 3NN
interaction can reproduce a weakly downwards dispersing state
which is relevant to superconductivity in graphene
nanoribbons~\cite{sasaki07-superconductivity}.

In this paper, we present a tight--binding (TB) formulation of the
$\pi$ bare energy band and QP dispersions of $AB$ stacked FLG and
graphite. We have previously compared both, $GW$ and LDA
calculations to ARPES experiments and proofed that LDA
underestimates the slope of the bands and trigonal
warping~\cite{partons06-graphite}. Here we list the TB fit
parameters of the QP dispersion (TB-$GW$) and the bare band
dispersion (TB-LDA) and show that the in--plane and out--of--plane
hopping parameter increase when going from LDA to $GW$. This new
and improved TB-$GW$ parameters is used for direct comparision to
experiments are obtained from a fit to QP calculations in the $GW$
approximation. This set of TB parameters works in the whole 2D
(3D) BZ of FLG (graphite) and is in agreement to recent
experiments. In addition we fit the parameters of the popular SWMC
Hamiltonian that is valid close to the $KH$ axis of graphite. This
paper is organized as follows: in section 2 we develop the 3NN TB
formulation for graphite and FLGs and in section 3 the SWMC
Hamiltonian is revised. In section 4 a new set of TB parameters
for the calculating the QP dispersion of stacked $sp^2$ carbon is
given. In section 5 we compare the graphite bare energy band (LDA)
to the QP ($GW$) dispersion. In section 6 we use the TB-$GW$
Hamiltonian and calculate the doping dependent Fermi surface of
graphite and estimate effective masses and free charge carrier
plasmon frequencies of pristine and doped graphite. In section 7
we show the calculated QP dispersions of FLGs. In section 8 we
discuss the present results and estimate the exciton binding
energies, transport properties and the low energy plasmon
frequencies. In section 9 the conclusions of this work are given.
Finally, in the appendices, the analytical forms of the
Hamiltonians for FLG are shown.

\begin{figure}
\begin{tabular}{c}
  \includegraphics[width=6cm]{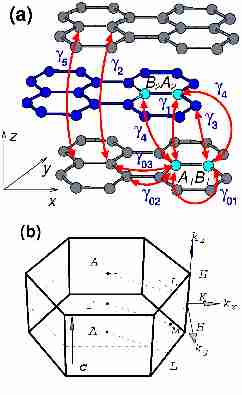}
\end{tabular}
\caption{(a) The graphite unit cell consists of four atoms denoted
by $A_1$,$B_1$,$A_2$ and $B_2$ (light blue). The red arrows denote
the interatomic tight--binding hopping matrix elements
$\gamma_0^1,\gamma_0^2,\gamma_0^3,\gamma_1,\ldots,\gamma_5$.
The overlap matrix elements $s_0-s_3$ (npot shown)
couple the same atoms as $\gamma_0^1-\gamma_0^3$.
(b) The 3D Brillouin zone of graphite with
the high symmetry points and the coordinate system used throughout
this work.} \label{fig:matrixel}
\end{figure}

\section{Third--nearest neighbour tight binding formulation}
Natural graphite occurs mainly with $AB$ stacking order and has
four atoms in the unit cell (two atoms for each graphene plane) as
shown in Fig.~\ref{fig:matrixel}(a). Each atom contributes one
electron to the four $\pi$ electronic energy bands in the 3D
Brillouin zone (BZ) [see Fig.~\ref{fig:matrixel}(b)]. FLG has $N$
parallel graphene planes stacked in an $AB$ fashion above one
another; the unit cell of FLG is 2D and the number of $\pi$ bands
in the 2D BZ equals $2N$. For graphite and FLG the TB calculations
are carried out with a new 3NN Hamiltonian and in addition with
the well--known SWMC Hamiltonian~\cite{rabi82-swmc} that is valid
in the vicinity to the Fermi level ($E_F$). The TB parameters that
enter these two Hamiltonians are
$\gamma=(\gamma_0^1,\gamma_0^2,\gamma_0^3,s_0,s_1,s_3,\gamma_1\ldots,\gamma_5,\Delta,E_0)$
for the 3NN Hamiltonian [shown in Fig.~\ref{fig:matrixel}(a)] and
$\gamma'=(\gamma'_0,\ldots,\gamma'_5,\Delta',E_0')$ for the SWMC
Hamiltonian. The hopping matrix elements for the SWMC Hamiltonian
are not shown here but they have a similar meaning with the
difference that only one nearest neighbour in--plane coupling
constant is considered (see e.g. Ref.~\cite{dresselhaus81} for an
explanation of SWMC parameters). The hopping matrix elements for
the 3NN Hamiltonian are shown in Fig.~\ref{fig:matrixel}(a). The
atoms in the 3D unit cell are labelled $A_1$,$B_1$ for the first
layer and $A_2$,$B_2$ for the second layer. The $A_2$ atom lies
directly above the $A_1$ atom in $z$ direction (perpendicular to
the layers). Within the $xy$ plane the interaction is described by
$\gamma_0^1$ (e.g. $A_1B_1$ and $A_2B_2$) for the nearest
neighbours, and  $\gamma_0^2$ and $\gamma_0^3$ for second nearest
and third nearest neighbours, respectively. A further parameter,
$\gamma_1$ ($A_1B_2$), is needed to couple the atoms directly
above each other (in $z$ direction). The hopping between adjacent
layers of sites that do not lie directly above each other is
described by $\gamma_3$ ($B_1B_2$) and $\gamma_4$ ($A_1B_2$ and
$B_1A_2$). The small coupling of atoms in the next--nearest layer
is $\gamma_2$ ($B_1B_1$ and $B_2B_2$) and $\gamma_5$ ($A_1A_1$ and
$A_2A_2$).

The calculation shown here is valid for both graphite and FLGs
with small adjustment as indicated when needed. The lattice
vectors for graphite in the $xy$ plane are ${\bf a_1}$ and ${\bf
a_2}$ and the out--of--plane lattice vector perpendicular to the
layers is ${\bf a}_3$.
\begin{equation}
\begin{array}{lll}
{\displaystyle {\bf a}_1=(\frac{\sqrt{3}a_0}{2},\frac{a_0}{2},0),
}& {\displaystyle {\bf
a}_2=(\frac{\sqrt{3}a_0}{2},-\frac{a_0}{2},0), }& {\displaystyle
{\bf a}_3=(0,0,2c_0). }
\end{array}
\label{eq:rab_nn}
\end{equation}
The C--C distance $\rm a_0=1.42\AA$ and the distance of two
graphene layers $\rm c_0=3.35\AA$. For a FLG with $N$ layers and
hence $2N$ atoms only the 2D unit vectors ${\bf a_1}$ and ${\bf
a_2}$. Similarly the electron wave vectors in graphite ${\bf
k}=(k_x,k_y,k_z)$  have 3 components and in FLG ${\bf
k}=(k_x,k_y)$. A TB method (or linear combination of atomic
orbitals, LCAO) is used to calculate the bare energy band and QP
dispersion by two different sets of interatomic hopping matrix
elements.  The electronic eigenfunction $\Psi({\bf r},{\bf k})$ is
made up from a linear combination of atomic $2p_z$ orbitals
$\phi({\bf r})$ which form the $\pi$ electronic bands in the
solid. The electron wave function for the band with index $\jmath$
is given by
\begin{equation}
\Psi^{\jmath}({\bf k},{\bf r})= \sum_{s={\rm A_1, B_1, \ldots
,B_N}}c^{\jmath}_s({\bf k}) \Phi_s({\bf k},{\bf r}), \ \ (\jmath =
1\ldots 2N). \label{eq:wavefun}
\end{equation}
Here $\jmath=1\ldots 2N$ is the electronic energy band index and
$s$ in the sum is taken over all atomic $2p_z$ orbitals from atoms
$A_1$,$B_1$,$A_2\ldots B_N$. Note that for 3D graphite we have
$N=2$. The $c^{\jmath}_s({\bf k})$ are wave function coefficients
for the Bloch functions $\Phi_s({\bf k},{\bf r})$. The Bloch wave
functions are given by a sum over the atomic wave functions
$\phi_s$ for each orbital in the unit cell with index
(${\ell},m,m$) multiplied by a phase factor. The Bloch function in
graphite for the atom with index $s$ is given by
\begin{equation}
\Phi_s({\bf r},{\bf
k})=\frac{1}{\sqrt{U}}\sum_{\ell,m,n}^U\exp(-i{\bf R}_s^{\ell
mn}\cdot {\bf k}) \phi_s({\bf r}-{\bf R}_s^{\ell mn}).
\end{equation}
where $U$ is the number of unit cells and $\phi_s$ denotes atomic
wave functions of orbital $s$. In graphite the atomic orbital
$\phi_s$ in the unit cell with index (${\ell},m,m$) is centered at
${\bf R}_s^{\ell mn}={\ell}{\bf a}_1+m{\bf a}_2+n{\bf a}_3+{\bf
r}_s$ and in FLGs at ${\bf R}_s^{\ell m}=\ell{\bf a}_1+m{\bf
a}_2+{\bf r}_s$.

For the case of FLGs $\Phi_s({\bf r},{\bf k}$) contains only the
sum over the 2D in--plane ${\bf R}_s^{\ell m}$. The $2N\times 2N$
Hamiltonian matrix defined by $H_{ss'}({\bf k})=\langle \Phi_s({\bf
r},{\bf k})|H({\bf r})|\Phi_{s'}({\bf r},{\bf k})\rangle$ and the
overlap matrix is defined by $S_{ss'}({\bf k})=\langle \Phi_s({\bf
r},{\bf k})|\Phi_{s'}({\bf r},{\bf k})\rangle$. For calculation of
$H$ and $S$, up to third nearest neighbour interactions (in
the $xy$ plane) and both nearest and next--nearest neighbour
planes (in $z$ direction) are included as shown in
Fig.~\ref{fig:matrixel}(a). The energy dispersion
relations are given by the eigenvalues $E({\bf k})$ and are
calculated by solving
\begin{equation}
H({\bf k})c({\bf k})=S({\bf k})E({\bf k})c({\bf k}).
\label{eq:hamil}
\end{equation}
In the Appendix, we show the explicit form of $H({\bf k})$ for
graphite FLGs with 1-3 layers.

\begingroup
\begin{table*}
\begin{tabular}{c|c|c|c|c|c|c|c|c|c|c|c|c|c|c}
\hline \hline
  Method & $\gamma_0^1$ & $\gamma_0^2$ & $\gamma_0^3$ &$s_0$ & $s_1$ & $s_2$ &$\gamma_1$ &$\gamma_2$ &$\gamma_3$ &$\gamma_4$ & $\gamma_5$ &  $E_0$ &  $\Delta$ \\
\hline\hline {\bf 3NN TB-${\bf GW}$} &  -3.4416 &
-0.7544&  -0.4246 & 0.2671 &  0.0494 &  0.0345 & 0.3513 & -0.0105
& 0.2973 &   0.1954 & 0.0187 &  -2.2624 & 0.0540\footnote{We
adjusted the impurity doping level in order to reproduce the
experimental value of $\Delta$.}\\
3NN TB-LDA &  -3.0121   & -0.6346 &  -0.3628 &   0.2499   & 0.0390 & 0.0322 &  0.3077 &  -0.0077 &   0.2583  &  0.1735 &  0.0147  & -1.9037&   0.0214 \\
\hline \hline
\end{tabular}
\caption{3NN tight--binding parameters for few--layer graphene and graphite.
The parameters of fits to LDA and $GW$ calculations are shown. The 3NN Hamiltonian is
valid in the whole two(three) dimensional BZ of graphite(graphene
layers).}\label{tab:gwval}
\end{table*}
\endgroup

\section{SWMC Hamiltonian}
The SWMC Hamiltonian has been extensively used in the
literature~\cite{dresselhaus81}. It considers only first nearest
neighbour hopping and is valid close to the $KH$ axis of graphite.
For small $k$ measured from the $KH$ axis (up to 0.15~$\rm\AA^{-1}$)
both, the 3NN and the SWMC Hamiltonians yield identical results.
The eight TB parameters for the SWMC Hamiltonian were previously
fitted to various optical and transport
experiments~\cite{dresselhaus81}.

For the cross sections of the electron and hole pockets analytical solutions
have been obtained and thus it has been used to calculate the
electronic transport properties of graphite.
Thus, in order to provide a connection to many transport experiments
from the past, we also fitted the LDA and $GW$ calculations to the
SWMC Hamiltonian.

The TB parameters are directly related to the energy band
structure. E.g. $\gamma_0$ is proportional to the Fermi velocity
in the $k_xk_y$ plane and $4\gamma_1$ gives the bandwidth in the
$k_z$ direction. The  bandwidth of a weakly dispersive band in
$k_z$ that crosses $E_F$ approximately halfway in between $K$ and
$H$ is equal to $2\gamma_2$, which is responsible for the semi
metallic character of graphite. Its sign is of great importance
for the location of the electron and hole pockets: a negative sign
brings the electron pocket to $K$ while a positive sign brings the
electron pocket to $H$. There has been positive signs of
$\gamma_2$ reported earlier~\cite{dresselhaus64-ibm} but it has
been found by M.S.~Dresselhaus~\cite{I9,dresselhaus81} that the
electron(hole) pockets are located at $K$($H$) which is in
agreement with recent DFT calculations~\cite{alex06-correlation},
tight--binding calculation~\cite{charlier91-graphite} and
experiments~\cite{dresselhaus81,lanzara06-graphite}. The magnitude
of $\gamma_2$ determines the overlap of electrons and holes and
the volume of the Fermi surface. It thus also strongly affects the
concentration of carriers and hence the conductivity and free
charge carrier plasmon frequency. The effective masses for
electrons and holes of the weakly dispersive energy band are
denoted by $m^*_{ze}$ and $m^*_{zh}$, respectively. Their huge
value also results from the small value of $\gamma_2$ and causes
the low electrical conductivity and low plasmon frequency in the
direction perpendicular to the graphene layers since
$m^*_{ze}$($\sqrt{m^*_{ze}}$) enters the denominator in the
expression for the Drude conductivity (free carrier plasmon
frequency). $\gamma_3$ determines the strength of the trigonal
warping effect ($\gamma_3=0$  gives isotropic equi--energy
contours) and $\gamma_4$ the asymmetry of the effective masses in
valence band (VB) and conduction band (CB). The other parameter
from next nearest neighbour coupling, $\gamma_5$ has less impact
on the electronic structure: both the VB and CB at $K$ are shifted
with respect to the Fermi level by $\Delta+\gamma_5$ causing a
small asymmetry~\cite{dresselhaus81}. Here $\Delta$ is the
difference in the on--site potentials at sites $A_1$($B_2$) and
$A_2$($B_1$). $\Delta$ is the value of the gap at the $H$
point~\cite{H8}. This crystal field effect for nonzero $\Delta$
occurs in $AB$ stacked graphite and FLGs but it does not occur in
$AA$ stacked graphite and the graphene monolayer. The small
on--site energy difference $\Delta$ appears in the diagonal
elements of $H({\bf k})$. It causes an opening of a gap at the $H$
point which results in a breakdown of Dirac Fermions in graphite
and FLGs with $N>1$. Finally, $E_0$ is set in such a way that the
electron and hole like Fermi surfaces of graphite yield an equal
number of free carriers. $E_0$ is measured from the bottom of the
CB to the Fermi level.

\begingroup
\squeezetable
\begin{table}
\begin{tabular}{c|c|c|c|c|c|c|c|c}
  \hline
\hline
  Method & $\gamma_0'$ &$\gamma_1'$ &$\gamma_2'$ &$\gamma_3'$ &$\gamma_4'$ & $\gamma_5'$ &  $E_0'$ &  $\Delta'$\\
\hline \hline
{\bf TB-${\bf GW}$}\footnote{This work}& 3.053&0.403&  -0.025 & 0.274 &  0.143&   0.030& -0.025& -0.005\footnote{We adjusted the impurity doping level to reproduce the experimental value of $\Delta'$}\\
TB-LDA$^a$&2.553&0.343&-0.018&0.180&0.173&0.018&-0.022&-0.018\\
\hline
EXP\footnote{Fit to Experiment, M.S.~Dresselhaus et al~\cite{dresselhaus81}}&3.16&0.39&-0.02&0.315&0.044&0.038&-0.024&-0.008\\
LDA\footnote{Fit to LDA, J.C.~Charlier et
al~\cite{charlier91-graphite}}&
2.598&0.364&-0.014&0.319&0.177&0.036&-0.026&-0.013\\
EXP\footnote{Fit to double resonance Raman spectra. L.M. Malard et al.\cite{pimenta07-bilayer}}& 2.9 & 0.3 & - & 0.1 & 0.12 & - & - \\
KKR\footnote{Fit to Korringa-Kohn-Rostocker first principles calculation\\ Tatar and Rabi~\cite{rabi82-swmc}}&2.92&0.27&-0.022&0.15&0.10&0.0063&0.0079&-0.027\\
 \hline \hline
\end{tabular}
\caption{The SWMC tight--binding parameters for the bare band dispersion
(LDA) and the quasiparticle dispersion ($GW$). All values are in eV.
This parameters are for the SWMC Hamiltonian
~\cite{dresselhaus81,rabi82-swmc,partons06-graphite,mikitik06-swmc}
which is valid close to the $KH$ axis. }\label{tab:fermivel}
\end{table}
\endgroup

\begin{figure}
\begin{tabular}{cc}
 \hspace{-0.7cm} \includegraphics[width=10cm]{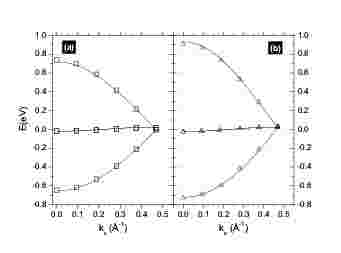}
\end{tabular}
\caption{The TB fit along $k_z$ direction at $k_z=k_y=0$ ($KH$
axis) for (a) LDA and (b) $GW$. $\Box$ denotes LDA calculations and
$\bigtriangleup$ denotes $GW$ calculations taken from Ref.~\cite{alex06-correlation} that were used for the
fitting. } \label{fig:kz}
\end{figure}
\begin{figure}
\begin{tabular}{cc}
 \hspace{-0.7cm} \includegraphics[width=10cm]{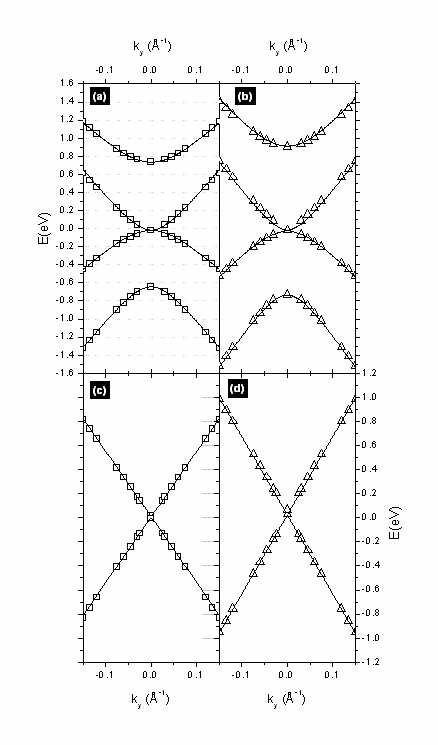}
\end{tabular}
\caption{Upper panel: The TB fits along $k_y$ direction at $\rm
k_z=0$ ($K$ point) for (a) LDA and (b) $GW$. Lower panel: the $k_y$
dispersion at $\rm k_z=0.47\AA^{-1}$ ($H$ point) for (c) LDA and
(d) $GW$. $\Box$ denotes LDA calculations and $\bigtriangleup$
denotes $GW$ calculations that were used for the fitting
(taken from Ref.~\cite{alex06-correlation}).}
\label{fig:ky}
\end{figure}
\begin{figure}
\begin{tabular}{cc}
  \hspace{-1cm}\includegraphics[width=10cm]{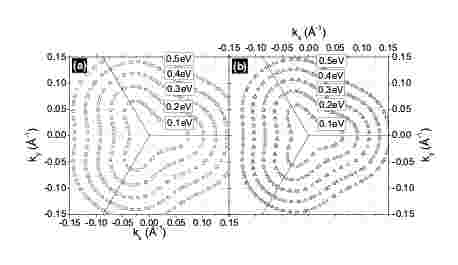}
\end{tabular}
\caption{The TB fits of equi--energy contours at $k_z$=0 for
(a) TB-LDA and (b)  TB-$GW$. } \label{fig:trig}
\end{figure}

\begin{figure*}
\begin{tabular}{cc}
    \includegraphics[width=7cm]{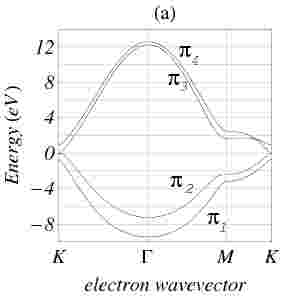}
  \includegraphics[width=7cm]{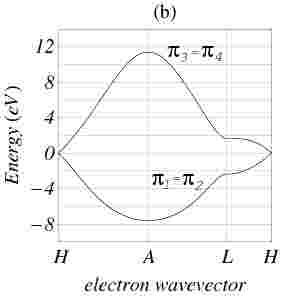}
\end{tabular}
\caption{The in--plane quasiparticle dispersion for (a) $k_z=0$
and (b) $k_z$=0.47~$\rm \AA^{-1}$ calculated by TB-$GW$. The symbols $\pi_1-\pi_4$ denote the four
$\pi$ bands of graphite.}
\label{fig:inplane}
\end{figure*}

\begin{figure}
\begin{tabular}{c}
\hspace{-1cm}\includegraphics[width=8cm]{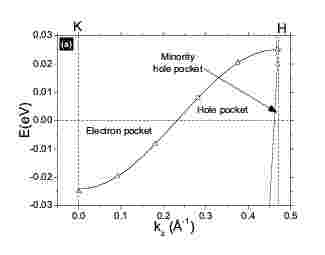}\\
\hspace{-1cm}\includegraphics[width=8cm]{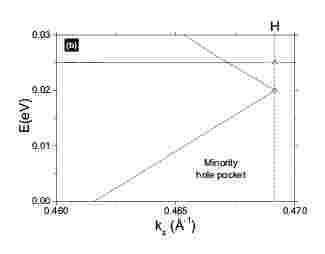}
\end{tabular}
\caption{(a) The weakly dispersive band that is responsible for the
formation of electron and hole pockets. $\bigtriangleup$ denotes
$GW$ calculations (with the impurity doping level adjusted to reproduce
the experimental gap at $H$ point) and the line the TB-$GW$ fits. (b) shows a magnification of
the minority hole pocket that is caused by the steeply dispersive band close to $H$.}
\label{fig:tblda_pockets}
\end{figure}

\begin{figure*}
\begin{tabular}{c}
\includegraphics[width=14cm]{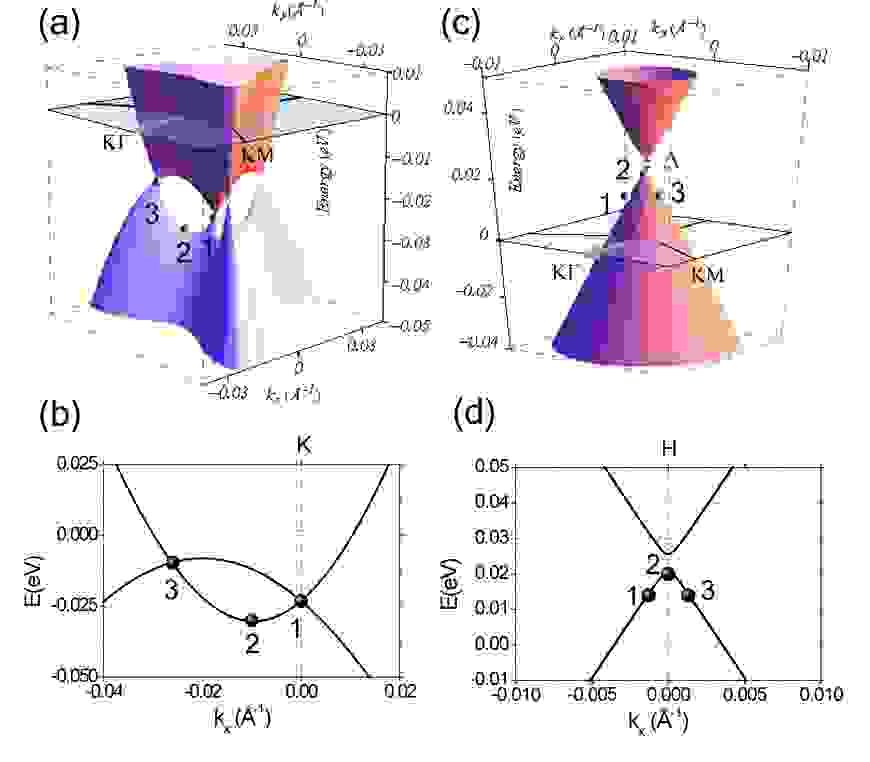}
\end{tabular}
\caption{TB-$GW$ QP band structure around (a,b) $K$ point, (c,d) $H$ point.
The points $1$,$2$ and $3$ denote the dispersion we use to determine electron and
hole masses (see Fig.~\ref{fig:mass}). The points $1$,$2$ and $3$
span a parabola. In (a,b) the parabola is along the $KM$ direction with heaviest electron
masses and in (c,d) the hole masses are isotropic around $H$. Note that the parabolas along $1-3$ are
used for evaluation of the effective electron and hole masses.}
\label{fig:pocket}
\end{figure*}

\begin{figure*}
\begin{tabular}{c}
\includegraphics[width=12.0cm]{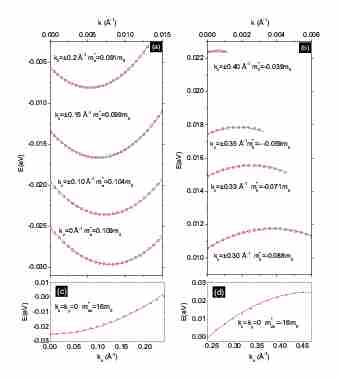}
\end{tabular}
\caption{Evaluation of the in--plane (a) electron and (b) hole
masses around $K$ and $H$, respectively. The dispersions along the
parabolas depicted by points $1-2-3$ in Fig.~\ref{fig:pocket} are
taken for evaluation of the masses. The effective masses
perpendicular to the layers in $z$ direction are shown in (c) for
electrons and in (d) for holes. $\Box$ correspond to TB-$GW$
values and the red lines are parabolic fits. } \label{fig:mass1}
\end{figure*}

\begin{figure}
\begin{tabular}{c}
\includegraphics[width=5.0cm]{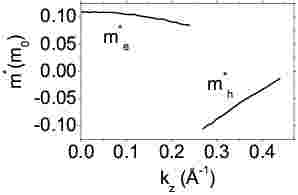}
\end{tabular}
\caption{The $k_z$ dependence of the in--plane electron mass ($m_e^*$) and hole
mass ($m_h^*$) calculated by TB-$GW$. The masses are evaluated along the
parabolas as shown in Fig.~\ref{fig:mass1}.
\label{fig:mass}}
\end{figure}

\section{Numerical fitting procedure}
The ab--initio calculations of the electronic dispersion are
performed on two levels: bare band dispersion calculation by LDA
and QP dispersion calculations within the $GW$ approximation. We
calculate the Kohn-Sham band-structure within the LDA to
density-functional theory (DFT)~\cite{abinit}. Wave-functions are
expanded in plane waves with an energy cutoff at 25 Ha. Core
electrons are accounted for by Trouiller-Martins pseudopotentials.

We then employ the $G_0W_0$ approximation using a plasmon-pole
approximation for the screening
\cite{hybert86,hedin65-gw,louie-gw} to calculate the self-energy
corrections to the LDA dispersion. For the calculation of the
dielectric function $\epsilon(\omega,q)$ we use a
15$\times$15$\times$5 Monkhorst-Pack $k$ sampling of the first BZ,
and conduction band states with energies up to 100~eV above the
valence band (80~bands), calculations were performed using the
code YAMBO~\cite{self}. The details for the first principles
calculations are given elsewhere~\cite{claudio08}.

For the fitting of the TB parameters to the ab--initio (LDA and $GW$) calculations we used energies of the four $\pi$ bands of graphite at
$\sim$~100 $k$ points. The points were distributed inside the
whole 3D BZ of graphite. The fitting was performed with the 3NN
Hamiltonian. In addition we chose a smaller subset of points
inside a volume of $\rm 0.15\AA^{-1}\times 0.15\AA^{-1} \times
0.47\AA^{-1}$ and fitted the SWMC parameters, which is frequently
used in the
literature~\cite{dresselhaus81,rabi82-swmc,partons06-graphite}).

The set of TB parameters were fitted by employing a
steepest--descent algorithm that minimizes the sum of squared
differences between the TB and the ab--initio calculations. This
involves solving Eq.~\ref{eq:hamil} with different sets of TB
parameters so as to approach a minimum deviation from the
ab--initio calculations. Points close to $E_F$ were given
additional weight so that the band crossing $E_F$ was described
with a deviation less than 1~meV. This is important for an
accurate description of the Fermi surface. In
Table\,\ref{tab:gwval} we list the parameters for the 3NN
Hamiltonian that can be used to calculate TB-$GW$ bands in the
whole 3D BZ of graphite.

The parameters that were fit with the SWMC Hamiltonian are
summarized in Table\,\ref{tab:fermivel}. These TB parameters
reproduce the bare energy band (fit to LDA) and the QP (fit to $GW$)
calculated dispersions. Hereafter these fits are referred to as
TB-LDA and TB-$GW$, respectively. We also list the values from other
groups that were fit to experiments~\cite{dresselhaus81} and to
LDA~\cite{charlier91-graphite} and another first--principles
calculation~\cite{rabi82-swmc}. It can be seen that the TB-$GW$
parameters for the nearest neighbour coupling increase by about
20\% when compared to TB-LDA. TB-$GW$ is also closer to the
experimental TB parameters than the TB-LDA parameters. This
indicates that electronic correlation effects play a crucial role
in graphite and FLGs for interpreting and understanding
experiments that probe the electronic energy band structure.

\section{Comparision of the bare energy band to the quasiparticle dispersion of graphite}

We now compare the calculated TB-LDA to TB-$GW$ and we also show the
result of the first--principles calculations that were used for
fitting in order to illustrate the quality of the fit. In
Fig.~\ref{fig:kz} the full $k_z$ dispersion from $K$ to $H$ for
(a) TB-LDA is compared to (b) TB-$GW$ calculations. It is clear that
the bandwidth in $k_z$ increases by about 20\% or 200~meV when
going from TB-LDA to TB-GW, i.e. when long--range
electron--electron interaction is taken into account. Such an
increase in bandwidth is reflected by the TB parameter $\gamma_1$
(in the SWMC model $4\gamma_1$ is the total bandwidth in the
out--of--plane direction). It can be seen that the conduction
bandwidth increases even more than the valence bandwidth. The VB
dispersion was measured directly by ARPES and gave a result in
good agreement to the TB-$GW$~\cite{alex06-correlation}.

The dispersion parallel to the layers is investigated in
Fig.~\ref{fig:ky} where we show the $k_y$ dispersion for $k_z=0$
($K$) and $k_z=0.47\rm\AA^{-1}$ ($H$). Here the TB-GW bands also
become steeper by about 20\% when compared to TB-LDA. This affects
$\gamma_0$ which is the in--plane nearest--neighbour coupling (see
Fig.~\ref{fig:matrixel})(a). It determines the in--plane $v_F$ and
in--plane bandwidth which is proportional to
$\gamma_0^1-\gamma_0^3$ (or proportional to $\gamma_0$ in the SWMC
Hamiltonian). In Fig.~\ref{fig:trig} the trigonal warping effect
is illustrated by plotting an equi--energy contour for $k_z=0$
with (a) TB-LDA and (b) TB-$GW$.
The trigonal warping effect is determined by $\gamma_3$
which is larger in the TB-$GW$ fit compared to the TB-LDA.

\section{The three dimensional quasiparticle dispersion and doping dependent Fermi surface of graphite}

\begin{table}
\begin{tabular}{c|c|c|c|c|c}
\hline
\hline
Point & Method & $\pi_1$ & $\pi_2$ & $\pi_3$ & $\pi_4$ \\
\hline
\hline
           & $GW$     & -9.458 & -7.257 & 12.176 & 12.541 \\
\rb{$\Gamma$}   & TB-$GW$  & -9.457 & -7.258 & 12.184 & 12.540 \\
\hline
      & $GW$     & -3.232 & -2.441 & 1.655 & 2.491 \\
\rb{$M$}   & TB-$GW$  & -3.216 & -2.457 & 1.656 & 2.495 \\
\hline
           & $GW$     &  -0.736 & -0.025 & -0.025 & 0.917 \\
\rb{$K$}        & TB-$GW$  &  -0.728 & -0.024 & -0.024 & 0.909 \\
\hline
           & $GW$     &  0.020 &  0.020 & 0.025 & 0.025 \\
\rb{$H$}        & TB-$GW$  &  0.020 &  0.020 & 0.025 & 0.025 \\
\hline
\end{tabular}
\caption{Energy values of the $GW$ calculation and the 3NN TB fit (TB-$GW$) at high symmetry points in the 3D BZ of graphite (all values in units eV).
The symbols $\pi_1$-$\pi_4$ denote the four $\pi$ bands of graphite.
The TB-$GW$ have been calculated with the 3NN TB-$GW$ parameters from Table~I.}\label{tab:vergleich}
\end{table}

In the previous section we have shown that the $\pi$ bandwidth
increases by 20\% when going from TB-LDA to TB-$GW$. This affects
especially the optical properties such as the
$\pi\rightarrow\pi^*$ transition that plays an important role in
optical absorption and resonance Raman and thus one has to use
TB-$GW$ for proper description of the electronic structure of
graphite including electron--electron correlation effects. In
Fig.~\ref{fig:inplane} we show the complete in--plane QP band
structure for (a) $k_z=0$ ($K$ point) and (b) $k_z=\rm
0.46~\AA^{-1}$ ($H$ point) calculated by the 3NN TB Hamiltonian.
In the $\Gamma KM$ plane two valence bands ($\pi_1$ and $\pi_2$)
and two conduction bands ($\pi_3$ and $\pi_4$) can be seen and in
the whole $HAL$ plane the two valence (conduction) bands are
degenerate. The ab-initio $GW$ values are compared to the 3NN
TB-$GW$ calculation in Table~\ref{tab:vergleich}. It can be seen
that the fit reproduces the ab--initio calculations with an
accuracy of $\sim $~10~meV in the BZ center and an accuracy of
1~meV along the $KH$ axis, close to $E_F$.

Close to $E_F$ the 3NN Hamiltonian is identical to the SWMC
Hamiltonian. Thus, for evaluation of the Fermi surface and the
doping dependence on $E_F$ in the dilute limit, we use the SWMC Hamiltonian.
The weakly dispersing band that crosses $E_F$ is
responsible for the Fermi surface and the electron
and hole pockets. This energy band is illustrated in
Fig.\ref{fig:tblda_pockets}(a) where we show that the TB-$GW$ fit has
an accuracy of 1~meV. The minority pocket that is a result of the
steeply dispersive energy band close to $H$ is shown in Fig.\ref{fig:tblda_pockets}(b).

The QP dispersion close to $E_F$ is shown in Fig.\ref{fig:pocket}
for (a,b) $K$, (c,d) $H$ point. The dispersion around the $K$
point is particular complicated: there are four touching points
between valence and conduction bands. Three touching points
between the valence and conduction bands exist in the close
vicinity to $K$ at angles of $\rm 0^o$, $\rm 120^o$ and $\rm
240^o$ away from $k_x$ (i.e. the $KM$ direction). The fourth
touching point is exactly at $K$ point. The touching points arise
from the semi metallic character of graphite: there are two
parabolas (VB and CB) that overlap by about 20~meV. For example
the bottom of the CB is denoted by the point $2$ in
Fig.\ref{fig:pocket}(a) and (b). At $H$ point shown in
Fig.\ref{fig:pocket}(c) and (d) the energy band structure becomes
simpler: there are only two non--degenerate energy bands and their
dispersion is rather isotropic around $H$ (the trigonal warping
effect in the $k_xk_y$ plane is a minimum in the $AHL$ plane and a
maximum in the $\Gamma KM$ plane). It is clear that the energy
bands do not touch each other and the dispersion is not linear but
parabolic with a very large curvature (and hence a very small
absolute value of the effective mass; see Fig.~\ref{fig:mass1}) at
$H$. $E_F$ lies 20~meV below the top of the VB and the energy gap
$\Delta$ is equal to 5~meV. It is interesting to note that the a
larger value of $\Delta$ would bring the top of the VB above
$E_F$. The horizontal cuts through the dispersions in
Fig.\ref{fig:pocket} at $E=E_F$ (blue area) give cross sectional
areas of an electron--like Fermi surface at $\rm k_z=0$ ($K$) and
a hole--like Fermi surface at $k_z=0.47~\rm\AA^{-1}$ ($H$ point),
consistent with a semi-metallic behaviour.

The effective in--plane massses for electrons ($m^*_e$) and holes
($m^*_h$) are evaluated along the parabola indicated in
Fig.\ref{fig:pocket} by the points $1,2,3$. This parabola lies in
the plane spanned by $KM$ and $HL$ and can thus be considered an
upper limit for the effective mass since the dispersion is flat
in this direction as can be seen in Fig.\ref{fig:pocket}(a). For
$m^*_e$ the center of the parabola is chosen to be the bottom of
the CB, i.e. point $2$ in Fig.\ref{fig:pocket}(a). Similarly for
$m^*_h$ the center of the parabola is chosen to be the top of the
VB, i.e. point $2$ in Fig.\ref{fig:pocket}(b). Due to the larger
curvature of the hole bands, the absolute value of $m^*_e$ is
larger than $m^*_h$. The $k_z$
dependence of $m^*_e$ and $m^*_h$ is shown in
Fig.\ref{fig:mass1}(a) and (b), respectively. For the effective mass in the $z$ direction we
fit a parabola for the weakly dispersing band in direction
perpendicular to the layers and get $m^*_{ze}=16m_0$ and
$m^*_{zh}=-16m_0$ for the effective electron and hole masses
perpendicular to the layers, respectively. The
Fig.\ref{fig:mass1}(c) and Fig.\ref{fig:mass1}(d) shows the TB-$GW$
along $k_z$ for the electron and hole pocket, respectively. The
parabolic fits that were used to determine the effective masses
are shown along with the calculation.

The $k_z$ dependence of $m_e^*$ and $m_h^*$ are shown in
Fig.~\ref{fig:mass}. For this purpose we evaluated the heavy
electron mass of the parabolic sub bands as shown in
Fig.~\ref{fig:pocket} and Fig.~\ref{fig:mass1}). It is clear that
$m_e^*$ has a weak $k_z$ dependence and $m_h^*$ strongly depends
on the value of $k_z$. This is obvious since exactly at $H$ point,
the value of $m_h^*$ has a minimum. For a finite value of the gap
$\Delta$, the value of $m_h^*$ also remains finite.

We now discuss the whole 3D Fermi surface. The volume inside the
surface determines the low--energy free carrier plasmon
frequencies and the electrical conductivity.  The trigonal warping
has little effect on the volume inside the electron and hole
pocket. When we set $\gamma_3=0$, then the Fermi surface is
isotropic around $KH$ axis. The simplification of $\gamma_3=0$
results in little change of the volume. For $\gamma_3\ne 0$, there
are touching points of the electron--like and hole--like Fermi
surfaces~\cite{dresselhaus81}. The touching points (or legs) are
important for understanding the period for de--Haas--van Alphen
and the large diamagnetism in graphite~\cite{mikitik06-swmc}.
However, for the calculation of the number of carriers, they are
not crucial and thus the Fermi surface calculated with
$\gamma_3=0$ can be used for the evaluation of the electron
density, $n_e$, and the hole density, $n_h$. In this case, the
cross section of the Fermi surface A($k_z$) has an analytical
form. The number of electrons per $\rm cm^3$ is given by
$n_e=4\times 10^{24}\times f_u/v_{uc}$ with $f_u=v_{e}/v_{bz}$
where $v_e$ is the electron pocket volume and $v_{bz}$ the BZ
volume. $v_{uc}$ is the unit cell volume in $\rm\AA^{-3}$.
Similarly, by replacing $v_{e}$ with $v_{h}$ (the volume of the
whole pocket) one can obtain $n_h$ the number of holes per $\rm
cm^3$. The critical quantities are $v_e$ and $v_h$ and they are
obtained by integrating the cross section of the Fermi surface,
$A(k_z)$ along $k_z$. The analytical expression and their
dependence on the TB parameters is given in~\cite{I9}. This yields
plasmon frequencies of
$\hbar\omega_a=\hbar\sqrt{n_ee^2/(m_e^*\epsilon_0\epsilon_a)}=\rm
113~meV$ for plasmon oscillation parallel to the graphene layers
and
$\hbar\omega_c=\hbar\sqrt{n_ee^2/(m^*_{ze}\epsilon_0\epsilon_c)}=\rm
19~meV$ for plasmon oscillation perpendicular to the layers. Here
$\epsilon_a=5.4$~\cite{venghaus75-epsilon} and
$\epsilon_c=1.25$~\cite{palmer91-hreels} are adopted for the
dielectric constants parallel and perpendicular to the graphene
layers, respectively. We have made two simplifications: first we
do not consider a finite value of temperature ($\omega_a$ and
$\omega_c$ are the plasmon frequencies at 0~K) and second we used
an effective mass averaged over the whole $k_z$ range of the
pockets as shown in Fig.\ref{fig:mass1} and in Fig.\ref{fig:mass}.

\begin{figure*}
\begin{tabular}{c}
\includegraphics[width=18cm]{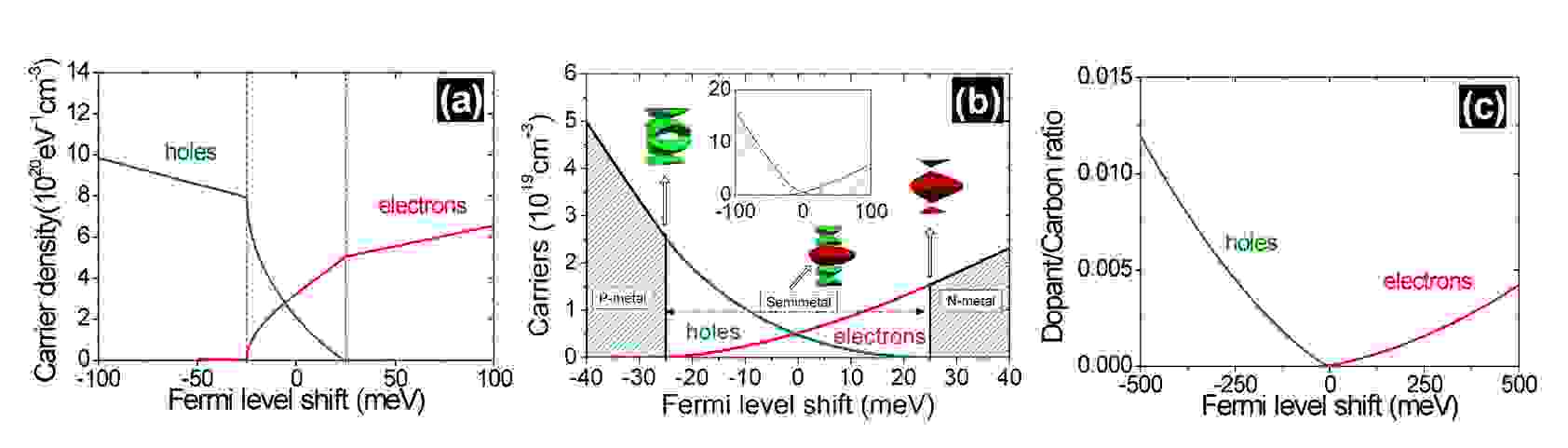}
\end{tabular}
\caption{Doping dependence of (a) the electron and hole carrier
density (b) the number of electrons and holes and (c) the
stochiometric dopant to carbon ratio. The red(green) lines
represent electron(hole) carriers.} \label{fig:doping}
\end{figure*}

\begin{figure*}
\begin{tabular}{c}
\includegraphics[width=17.5cm]{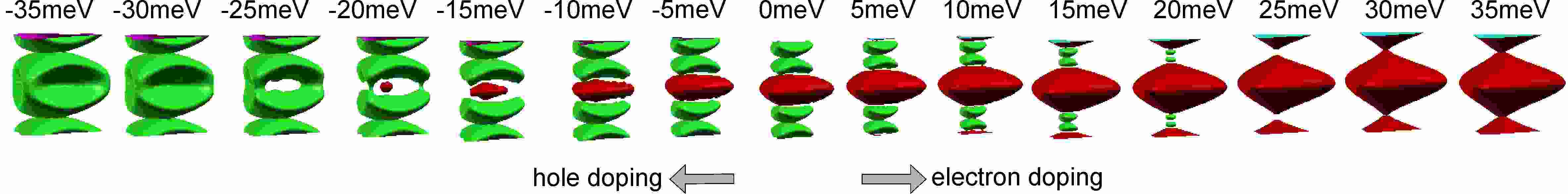}\\
\end{tabular}
\caption{The Fermi surface of doped graphite in the dilute limit
for electron doping and hole doping. The red(green) surfaces are
the electron(hole) pockets of the Fermi surface. It is clear that
at $\pm 25$~meV we have a semi metal to metal transition.}.
\label{fig:ehdoping_series}
\end{figure*}

Next we discuss the doping dependence of the electronic properties
in the so--called dilute limit, which refers to a very low ratio
of dopant/carbon atoms. Here we use a method described
previously~\cite{I9} employing the TB-$GW$ parameters from
Table~\ref{tab:fermivel}. This allows us to calculate the doping
dependence of $n_e$ and $n_h$. The analytical formula for the
$E_F$ dependent cross section of the Fermi surface, $A(k_z)$
(given in~\cite{I9}) is integrated for different values of $E_F$.
By integrating $dA(k_z)/dE_F$ along $k_z$ we obtain the carrier
density per eV. The ratio of dopant to carbon is given by
$r=n\times v_{uc}/4f$ where $n=n_e$ for electron doping and
$n=n_h$ for hole doping. Here $f$ is the charge transfer value per
dopant atom to Carbon. Although there are some discussions about
the value of $f$, it is was recently found for potassium doping
that $f=1$~\cite{alex08-kc8}. In Fig.~\ref{fig:doping}(a) we show
the doping dependence of the carrier densities. It is clear that
at $E_F=\pm 25~\rm meV$, we have a discontinuity in the carrier
density and this is associated to the $E_F$ at which the electron
or hole pocket is completely filled. Since the density of states
decreases suddenly after the pockets are filled, the kink in the
density of states appears. This also marks the transition from a
two--carrier regime to a single--carrier regime. In
Fig.~\ref{fig:doping}(b) we show $n_e$ and $n_h$. It is clear that
at $E_F=0$ the number of holes equals the number of electrons. At
$E_F=25~\rm meV$, we have no more holes and thus a transition from
a semi metal to a metal occurs. In such a metal, the carriers are
electrons, hence an N--type metal. Similarly, at $E_F=-25$~meV we
have no more electrons and the a semi metal to metal transition
occurs in the other direction to $E_F$. For this metal, the
carriers are holes, hence a P--type metal. These semi metal to
metal transitions are important for ambipolar transport in
graphite and graphene: they determine the region for the gate
voltage in which ambipolar transport is possible. Finally in
Fig.~\ref{fig:doping}(c) we plot the ratio of dopant to carbon
atoms as a function of $E_F$.

It is certainly interesting to monitor the doping induced changes
in $n_e$ and $n_h$ also in the shape of the Fermi surface. In
Fig.\ref{fig:ehdoping_series} we show the Fermi surfaces for
electron doping and hole doping for $E_F=0$ to $\pm 35~meV$ in steps of 5~meV. It is clear that at
$\pm 25~\rm meV$ a single carrier regime dominates as indicated by
the two different colors (red for electrons and green for holes).
This is consistent with Fig.~\ref{fig:doping}(b) where the
integrated electron (hole) densities disappear at
$E_F=25\rm~meV$ ($E_F=-25\rm~meV$).

\begingroup
\begin{table*}
\begin{tabular}{c|c|c|c}
  \hline
\hline
  Parameter & Symbol & TB-$GW$ & Experimental value(s)   \\
  \hline
  Fermi velocity at $H$ [$10^6$ms$^{-1}$] & $v_F$ & 1.01  & 0.91~\cite{lanzara06a-graphite}, 1.06~\cite{alex06-correlation}, 1.07~\cite{andrei07-ll}, 1.02~\cite{orlita08-gap} \\
  Splitting of $\pi$ bands at $K$ [eV] & $\delta$  &  0.704   & 0.71~\cite{alex06-correlation} \\
 Bottom of $\pi$ band at $A$ point [eV] & E($A$) & 7.6 & 8~\cite{takahashi06-prl}, 8~\cite{law86-graphite}\\
in--plane electron mass [$m_0$] &  $m_e^*$ & 0.1 ($k_z$ averaged) & 0.084~\cite{mendez79-effective_massa}, 0.42~\cite{lanzara06a-graphite},0.028~\cite{andrei07-ll}   \\
in--plane hole mass [$m_0$]\footnote{To compare different notations, we denote here
the absolute value of $m_h^*$.} & $m_h^*$ & 0.06 ($k_z$ averaged) &
0.069~\cite{lanzara06a-graphite},
0.03~\cite{Galt56-mass}\footnote{Mass at the $H$ point was
measured. Note that this is in excellent agreement to our calculated $k_z$ dependence of $m_h^*$ (see Fig.~9).}, 0.028~\cite{andrei07-ll} \\
  out--of--plane electron mass [$m_0$] & $m_{ze}^*$ & 16 & -   \\
  out--of--plane hole mass [$m_0$] & $m_{zh}^*$ & -16 &  - \\
  number of electrons at $E_F$=0 [$10^{18}$~cm$^{-3}$] & $n_e$ & 5.0 & 8.0~\cite{lanzara06a-graphite}, 3.1~\cite{soule58-mass}    \\
  number of holes at $E_F$=0 [$10^{18}$~cm$^{-3}$] & $n_h$ & 5.0 &  3.1~\cite{lanzara06a-graphite}, 2.7~\cite{soule58-mass}, 9.2~\cite{takahashi06-prl}  \\
  Gap at $H$ point [meV] & $\Delta$ & 5 & 5-8~\cite{orlita08-gap,H8}  \\
  in--plane plasmon frequency [meV]  & $\hbar\omega_a$ & 113 & 128~\cite{geiger71-plasmon}  \\
  out--of--plane plasmon frequency [meV] & $\hbar\omega_c$ & 19 &
  45-50~\cite{palmer91-hreels,geiger71-plasmon,palmer96-damping1}\\
  $A_1$ optical transition energy [eV] & $E_{A1}$ & 0.669 &  0.722~\cite{J46}\footnote{The difference between the experimental and the TB-$GW$ value might be a result of excitonic effects.} \\
  $A_2$ optical transition energy [eV] & $E_{A2}$ & 0.847 &  0.926~\cite{J46}$^{b}$ \\
  \hline
  \hline
\end{tabular}
\caption{Properties of the electronic band structure of graphite
calculated from TB-$GW$ and compared to
experiment.\label{table:prop}}
\end{table*}
\endgroup

\section{Few layer graphene\label{sec:flg}}
\begin{figure*}
\includegraphics[width=12cm]{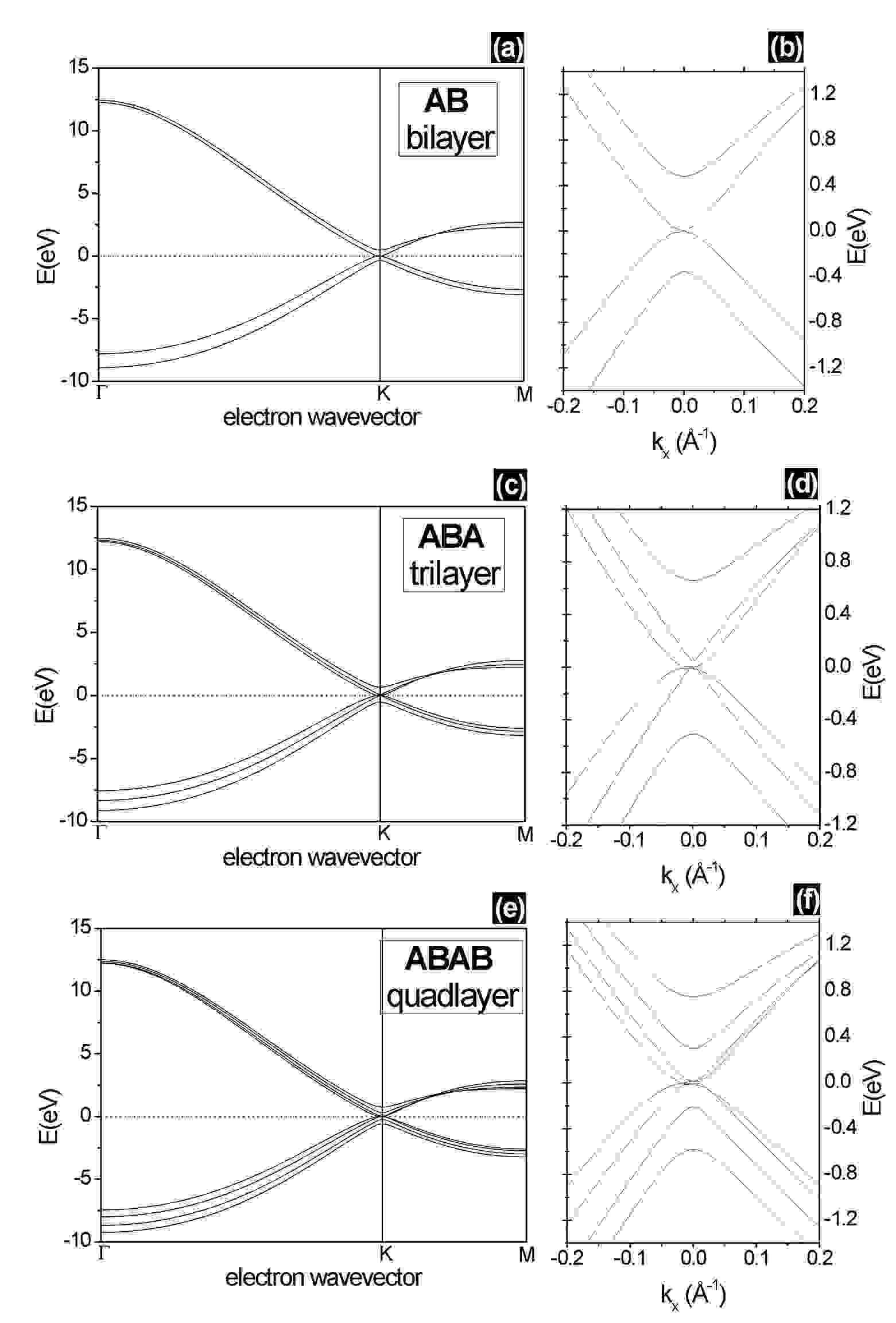}
\caption{Band structure around $K$ point of a bi-, tri- and
quadlayer calculated by TB-$GW$.} \label{fig:bilayer}
\end{figure*}

\begin{figure}
\includegraphics[width=9cm]{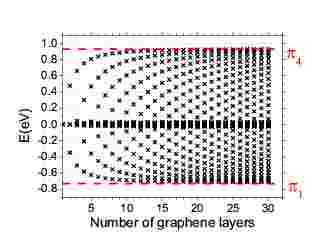}
\caption{Evolution of the eigenvalue spectrum for FLG from
$N=1\ldots 30$ graphene layers calculated by TB-$GW$ at the $K$
point. The dashed lines labelled by $\pi_1$ and $\pi_4$ denote the
lower and upper limits of the total bandwidth at $K$ point in 3D
graphite. A pattern that connects the first, second etc. energies
of FLG is emerging (see text).} \label{fig:flgtographite}
\end{figure}

The 3NN TB-$GW$ set of parameters fits the whole $k_z$ range of
the 3D graphite BZ. Thus the set can be transferred for the
calculation of QP dispersions of stacked $sp^2$ FLGs with $N$
layers ($N=1,2,\dots$). The transferability of TB parameters is a
result of the fact that the lattice parameters of FLGs and
graphite are almost identical~\cite{trickey92-bilayer}. We can use
the matrix elements shown in Fig.~\ref{fig:matrixel}(a) also for
FLGa; for $N=1$ only $\gamma_0^1-\gamma_0^3$, $s_0-s_3$ and $E_0$
are needed and this results in the graphene monolayer case. For
$N=2$ the parameters $\gamma_2$ and $\gamma_5$ are not needed
since they describe next nearest neighbour interactions which do
not exist in a bilayer. We use the set of TB parameters given in
Table~\ref{tab:gwval} and the Hamiltonians given in section 2
and the appendix. In Fig.~\ref{fig:bilayer} we show the bilayer
($N=2$), the trilayer ($N=3$) and the quadlayer ($N=4$) calculated
with TB-$GW$. The Fig.~\ref{fig:bilayer}(a) and (b) shows the
electron dispersion of the bilayer. The separation between CB (VB)
to the Fermi level is proportional to
$\gamma_1$~\cite{mccann06-bilayer} and thus it is clear that the
TB-$GW$ also gives an about 20\% larger separation between the VB
than LDA. Furthermore the slope of all bands becomes steeper for
TB-$GW$ since $\gamma_0^1-\gamma_0^3$ increase with respect to LDA
calculations. This is also responsible for the increase in $v_F$
of FLGs (similar to the graphite case, when going from LDA to
$GW$). The same argument is the case for the tri-- and the
quadlayer. It is interesting that the QP
dispersion measured by ARPES~\cite{rotenberg06-graphite_bilayer}
are in better agreement with the TB-$GW$ rather than the LDA
calculations performed.

The low energy dispersion relation of FLGs are particularly
important for describing transport properties. From the
calculations, we find that all FLG has a finite density of
states at $E_F$. The trilayer has a small overlap at $K$ point between
valence and conduction band (i.e. semi--metallic). This property might be useful for devices with
ambipolar transport properties. Our results are in qualitative
agreement with the LDA calculations~\cite{henrard06-hpoint}.

It can be seen that a linear (Dirac--like) band appears for the
trilayer (and also for all other odd-numbered multilayers).
This observation is in agreement to previous calculations and is relevant to a increase
in the orbital contribution to diamagnetism~\cite{ando08-magnetism}.

Since TB allows for rapid calculation of the QP bands, the
transition from FLG to bulk graphite can be analyzed. Even in the
case of $N=30$, the solution of Eq.~\ref{eq:hamil} for the
$60\times 60$ Hamiltonian and overlap matrices takes only
$\sim$~10 sec on a Pentium~III workstation per $k$ point. In
Fig.~\ref{fig:flgtographite} we show the eigenvalue spectrum for
FLG with $N=1,\ldots,30$. As we increase $N$ and hence the number
of $\pi$ bands, the bandwidth also increases and approaches that
of bulk graphite. It can be seen that for $N>15$, the total
bandwidth is that of bulk graphite. Interestingly, the energies of
the $\pi$ bands group together and form families of the highest,
second highest etc. energy eigenvalue at $K$. With increasing
number of layers, a given family approaches the limit for bulk
graphite. Such a family pattern is a direct consequence of the
$AB$ stacking sequence in FLG and it might be accessible to
optical spectroscopy similar to the fine structure around $K$
point that has been observed in bulk graphite~\cite{J46}.

\section{Discussion}
We first discuss the QP band structure and relation to recent
ARPES experiments. From several experimental works it is clear
that the LDA bands need to be scaled in order to fit the
experiments~\cite{rotenberg06-graphite,zhou05-graphite,alex06-correlation}.
Our new set of TB parameters quantitatively describes the QP
dispersions of graphite and FLG. The scaling is mainly reflected
in an increase of $\gamma_0^1-\gamma_0^3$ and $\gamma_1$, the
in--plane and out--of--plane coupling, respectively. It thus can
be used to analyze ARPES of both pristine and doped (dilute limit)
graphite and FLG.  Most importantly, the correlation effects
increase $v_F$, the Fermi velocity when going from LDA to $GW$.
For the TB-$GW$ calculation, $v_F=1.01\times 10^6$ms$^{-1}$ which
is in perfect agreement to the values from ARPES that is equal to
$v_F=1.06\times 10^6$ms$^{-1}$~\cite{alex06-correlation}.
The question why LDA works for some metals but fails to give the
correct $v_F$ and energy band dispersion in semi metallic graphite
arises. In graphite, the contribution of the electron--electron
interaction to the self--energy is unusually large. The reason for
this is the small number of free carriers to screen efficiently
the Coulomb interaction. In most other metals, the density of
states at $E_F$ has a much larger ($\sim$ 1000 times) value than
in graphite and the screening lengths are shorter. Hence the LDA
is a good description for such a material but it fails in the case
of graphite.
Another parameter that illustrates the quality of the TB-$GW$ to
reproduce the experimental QP dispersion is $\delta$, the band splitting at $K$ point.
For TB-$GW$, we obtain $\delta=0.704$~eV and the ARPES gives $\delta=0.71$~eV~\cite{alex06-correlation}.

The TB parameters of the SWMC Hamiltonian have been fitted in
order to reproduce double--resonance Raman
spectra~\cite{pimenta07-bilayer}. As a result they obtained that
$\gamma_3$ (one of the parameters that couples the neighbouring
graphene planes) has a value of 0.1~eV while the for graphite that
we obtain in this paper is 0.274~eV ´(see Table II) and the value
that was fit to transport experiments is
0.315~eV~\cite{alex07-kirchberg,dresselhaus81} which is also in
perfect agreement to ARPES experiments of graphite single
crystals~\cite{alex07-kirchberg}. We now discuss a possible reason
for this discrepancy. The fitting procedure in
Ref~\cite{pimenta07-bilayer} depends on the choice of the phonon
dispersion relation of graphite. It is important to note that
recently a Kohn anomaly has been directly observed by inelastic
x--ray scattering experiments using synchrotron
radiation~\cite{alex08-ixs} which has a steeper slope of the TO
phonon branch at $K$ point in contrast to previous
measurements~\cite{maultzsch04}. Thus the assumption of the
correct phonon dispersion relation is crucial for obtaining the
correct band structure parameters.

Next we discuss the present QP electronic energy band structure in
relation to optical spectroscopies such as optical absorption
spectroscopy (OAS) and resonance Raman spectroscopy. The optical
spectroscopies probe the joint density of states (JDOS) weighted
with the dipole matrix elements. Peaks in the OAS are redshifted
when compared to the JDOS of the QP dispersion if excitons are
created. By comparing the QP dispersion with OAS
experiments~\cite{J46,e41,zhang87} we now estimate a value for
exciton binding energies in graphite. In general many resonant
states with $k$ contribute to OAS but a fine structure in OAS of
bulk graphite measured in reflection geometry was observed by Misu
et al.~\cite{J46} which was assigned to two specific transitions
around $K$: the $A_1$ transition between the lower VB and states
just above $E_F$ and the $A_2$ transition between the upper VB and
the upper CB. The experimental energies they found were
$E^{exp}_{A1}=0.669$~eV and $E^{exp}_{A2}=0.847$~eV. The TB-GW
dispersion yields energies $E^{GW}_{A1}=0.722$~eV and
$E^{GW}_{A2}=0.926$~eV [see Fig.\ref{fig:kz}(b)]. Assuming one
exciton is created for the $A1$($A2$) transition, this yields
exciton binding energies of $E^{GW}_{A1}-E^{exp}_{A1}\rm\sim
50$~meV and $E^{GW}_{A2}-E^{exp}_{A2}\rm\sim 80$~meV for a $K$
point exciton in bulk graphite. Such a value for the exciton
binding energies most probably increases when going from bulk
graphite to FLGs due to confinement of the exciton wave function
in $z$ direction.

Concerning the value of $\Delta$, the gap at $H$, several
experimental values exist and magneto reflectance experiments
suggest $\Delta$=5~meV. This is in disagreement to the calculated
values obtained for pristine graphite~\cite{alex06-correlation}.
However, when the doping level is slightly increased, $\Delta$
becomes smaller and we thus fixed a doping level in the ab--initio
calculation that reproduces the experimental
$\Delta$~\cite{claudio08}. While some of the variations may be
explained by the sample crystallinity in $z$ direction, it is also
conceivable that small impurities are responsible for the
discrepancy.

Next we discuss a possibility to measure the free carrier plasmon
frequencies of the of pristine and alkali--metal doped graphite by
high resolution energy electron loss spectroscopy (HREELS). The
electron concentration inside the pockets very sensitively affects
the plasmon frequencies. In principle the charge carrier plasmons
should also appear as a dip in optical reflectivity measurements
but due to the small relative change in intensity they have not
been observed so far. Due to the small size of the pockets, the
number of charge carriers and hence the conductivity and
plasmon frequencies are extremely sensitive to temperature and
doping. Experimentally observed plasmon frequencies for
oscillations parallel to the graphene layers are
$\hbar\omega_a$=$\rm 128~meV$~\cite{geiger71-plasmon} and for
oscillations perpendicular to the layers are $\hbar\omega_c$=$\rm
45-50~meV$~\cite{palmer91-hreels,geiger71-plasmon,palmer96-damping1}.
These values for $\hbar\omega_a$ and $\hbar\omega_c$ agree
reasonably well with our derived value considering that we make
the crude estimation of $\hbar\omega_a$ and $\hbar\omega_c$ at 0~K
while experiments where carried out at room temperature. We also
used average effective masses of the electron pocket in order to
determine the plasmon frequency, while in fact there is a $k_z$
dependence (see Fig.\ref{fig:mass}) . We also note that the experimental literature values
for $\epsilon_a$ and $\epsilon_c$  have a rather wide range, e.g.
$\epsilon_c=3.4$~\cite{zanini77-epsilon} from reflectivity
measurements and $\epsilon_c=5.4$~\cite{venghaus75-epsilon} from
EELS.

The temperature dependence of $\omega_c$ was measured by Jensen et
al \cite{palmer91-hreels} by HREELS and they observed a strong T
dependence for $\omega_c$ which was attributed to changes in the
occupation and thus number of free carriers with T. The observed
plasmon energy was rising from 40~meV to 100~meV in a temperature
range of 100~K to 400~K. A similar effect might be observed by
HREELS of doped graphite as a function of doping level. It would
be interesting to study the evolution of the plasmon frequency
with doping level. With our current understanding of the
low--energy band structure we predict a semimetal to metal
transition at a Fermi level shift of $E_F\sim 25$~meV. At this
doping level, the hole pocket is completely filled with electrons
and disappears and the electron pocket has roughly doubled in size
and $\omega_a$ and $\omega_c$ should increase by a factor of
$\sim\sqrt{2}$.  Such a transition should be observable by HREELS
and might be accompied by interesting changes in the band
structure (electron--plasmon coupling) which can also be measured
simultaneously by ARPES. Many DC transport properties can be
understood with a Drude model for the conductivity, which is
inversely proportional to the effective carrier mass. From the
ratio of $m^*_{ze}/m^*_e$ and $m^*_{zh}/m^*_h$, we hence expect
that the DC electron(hole) conductivity in $z$ direction is $\sim
200$($\sim 500$) times less than the in--plane conductivity . The
experimental value is $\rm 3\times 10^3$\cite{dresselhaus81} and
considering the large variation in experimental values even for
the in--plane
conductivity~\cite{morelli84-resistivity,dresselhaus81}, it is in
reasonable agreement. In Table~\ref{table:prop} we summarize the
electronic band structure properties such as $v_F$, $\delta$, the
effective masses etc. and compare them to experimental values.

\section{Conclusions}
In conclusion, we have fitted two sets of TB parameters to first
principles calculations for the bare band (LDA) and QP ($GW$)
dispersions of graphite. We have observed a 20\% increase of the
nearest neighbour in--plane and out--of--plane matrix elements
when going from LDA to $GW$. Comparision to ARPES and transport
measurements suggest that LDA is not a good description of the electronic band structure
of graphite because of the importance of correlation effects.
We have explicitely shown that the accuracy of the TB-$GW$ parameters is sufficient to reproduce the
QP band structure with an accuracy of $\sim$~10~meV for
the higher energy points and $\sim$~1~meV for the bands that are relevant for electronic transport.
Thus the 3NN TB-$GW$ calculated QP band energy dispersions are sufficiently accurate to be compared to a
wide range of optical and transport experiments. The TB-$GW$ parameters from Table~I should be used in the future
for interpretation of experiments that probe the band structure in graphite and FLG.
For transport experiments that only involve electronic states close to the $KH$ axis (in graphite) or the $K$ point in
FLG, the SWMC Hamiltonian with parameters from Table~II can be used.

With the new set of TB parameters we have calculated the low energy properties of graphite: the Fermi velocity, the
Fermi surface, plasmon frequencies and the ratio of conductivities
parallel and perpendicular to the graphene layers and effective
masses.  These values are compared to experiments and we have demonstrated an excellent agreement
with almost all values.

We have shown that at $\pm 25$~meV, a semi metal to metal
transition exists in graphite. Such a transition should be
experimentally observable by HREELS or ARPES for electron doped
graphite single crystals with very low doping levels (i.e. the
dilute limit). Experimentally, the synthesis of such samples is
possible by evaporation of potassium onto the sample surface and
choosing a sufficiently high equilibration temperature so that no
staging compound forms. It would thus be interesting to do a
combined ARPES and HREELS experiment on this kind of sample.

We have also discussed the issue of the existence of Dirac Fermions in graphite
and FLGs. For pristine graphite it is
clear that for a finite value of the gap $\Delta$ at $H$ point, we always have
a non--zero value of the effective masses at $H$. We now discuss some possibilities for $\Delta=0$. One way would be to use $AA$ stacked graphite. However, natural single crystal graphite has mainly $AB$ stacking and the preparation of
an $AA$ like surface by repeated cleaving is a rather difficult task. Recently, the possibility of reducing $\Delta$ by increasing the doping level has been put forward~\cite{alex08-kc8,claudio08} and this seems a more promising way for experiments. For the case of a fully doped graphite single crystal, the
interlayer interaction is negligible because the distance between the graphene
sheets is increased and the stacking sequence is change to $AA$ stacking. These two effects would cause the appearance of Dirac Fermions in doped graphite.

With the new set of 3NN TB parameters we have calculated the QP dispersion of FLG in
the whole 2D BZ. We have shown the evolution of the energy eigenvalue spectrum at $K$ point as a function of $N$, the number of
layers. An interesting family pattern was observed, where the first,second, third etc. highest transitions form a pattern that
approaches the energy band width of bulk graphite with increasing $N$. For the highest transition, the bulk graphite band width is already
reached at $N=15$. This family pattern is a direct result of the $AB$ stacking sequence in FLG and it can possibly be observe
experimentally. It should be pointed out here, that a very similar fine structure around the $K$ point in bulk graphite has been observed by optical
absorption spectroscopy by Misu et al.~\cite{J46} although the family pattern from Fig.~\ref{fig:flgtographite} might be observed also
by ARPES on high quality FLGs synthesized on Ni(111).

\section*{Acknowledgement}
A.G. acknowledges a Marie Curie Individual Fellowship (COMTRANS)
from the European Union. A.R. and C.A. are supported in part
by Spanish MEC (FIS2007-65702-C02-01), Grupos Consolidados UPV/EHU
of the Basque Country Government (IT-319 -07) European Community
e-I3 ETSF and SANES (NMP4-CT-2006-017310) projects. L.W.
acknowledges support from the French national research agency.
R.S. acknowledges MEXT grants Nos. 20241023 and 16076201.
We acknowledge M.~Knupfer for critical reading of the manuscript.
XCrysDen~\cite{xcrysden} was used to prepare Fig.~10.

\newpage
\section*{Appendix}
We made computer programs to automatically generate the 3NN Hamiltonians
and overlap matrices for $N$ graphene layers. The method is following
the description by Partoens et al.~\cite{partons06-graphite} but with the
difference that we include 3NN matrix elements that describe the QP dispersion in the whole
BZ. It is interesting to note the similarity between the Hamiltonian of a $N$ layer graphene
and a stage $N$ graphite intercalation compound~\cite{saito86-gic}.
The general form of the Hamiltonian is given by
\begin{eqnarray*}
\nonumber {\displaystyle H({\bf k})=\left (
\begin{tabular}{ccccc}
$H_{A_1A_1}$&$H_{A_1B_1}$&$H_{A_1A_2}$&\ldots&$H_{A_1B_N}$\\
$H_{B_1A_1}$&$H_{B_1B_1}$&$H_{B_1A_2}$&\ldots&$H_{B_2B_N}$\\
$H_{A_2A_1}$&$H_{A_2B_1}$&$H_{A_2A_2}$&\ldots&$H_{A_2B_N}$\\
\vdots&\vdots&\vdots&\vdots&\vdots\\
$H_{B_NA_1}$&$H_{B_NB_1}$&$H_{B_NA_2}$&\ldots&$H_{B_NB_N}$
\end{tabular}
\right )}
\end{eqnarray*}
and the overlap matrix is given by
\begin{eqnarray*}
\nonumber {\displaystyle S({\bf k})=\left (
\begin{tabular}{ccccc}
$S_{A_1A_1}$&$S_{A_1B_1}$&$S_{A_1A_2}$&\ldots&$S_{A_1B_N}$\\
$S_{B_1A_1}$&$S_{B_1B_1}$&$S_{B_1A_2}$&\ldots&$S_{B_2B_N}$\\
$S_{A_2A_1}$&$S_{A_2B_1}$&$S_{A_2A_2}$&\ldots&$S_{A_2B_N}$\\
\vdots&\vdots&\vdots&\vdots&\vdots\\
$S_{B_NA_1}$&$S_{B_NB_1}$&$S_{B_NA_2}$&\ldots&$S_{B_NB_N}$
\end{tabular}
\right ).}
\end{eqnarray*}

The QP band structure is then given by solving equation Eq.~\ref{eq:hamil}.
The 3NN TB Hamiltonians can be used with $GW$ parameters from
Table~\ref{tab:gwval}. In the following we give the explicit form
for the $H({\bf k})$ and $S({\bf k})$ of a monolayer and an $AB$ stacked bi-- and and $ABA$ trilayer which was used in section~\ref{sec:flg}.

\subsection*{Monolayer graphene}
The monolayer 3NN Hamiltonian and overlap matrices are given by

\begin{eqnarray*}
\nonumber {\displaystyle H({\bf k})=} {\displaystyle \left (
\begin{tabular}{cc}
\e0+$\gamma_0^2$\effzwei&$\gamma_0^1$\eff+$\gamma_0^3$\effdrei\\
$\gamma_0^1$\effs+$\gamma_0^3$\effsdrei&\e0+$\gamma_0^2$\effszwei
\end{tabular}
\right )}
\label{eq:s}
\end{eqnarray*}
and
\begin{eqnarray*}
\nonumber {\displaystyle S({\bf k})=} {\displaystyle \left (
\begin{tabular}{cc}
1+\effzwei$s_0^2$&\eff$s_0^1$+\effdrei$s_0^3$\\
\effs$s_0^1$+\effsdrei$s_0^3$&1+\effszwei$s_0^2$
\end{tabular}
\right )} .\label{eq:s}
\end{eqnarray*}

\newpage
\subsection*{Bilayer graphene}

The bilayer 3NN Hamiltonian and overlap matrices are given by
\begin{widetext}
\begin{eqnarray}
\nonumber {\displaystyle H({\bf k})=}\\ {\displaystyle \left (
\begin{tabular}{cccc}
\enulld+$\gamma_0^2$\effzwei& $\gamma_0^1$\eff+$\gamma_0^3$\effdrei &$\gamma_1$&$\gamma_4$\effs\\
$\gamma_0^1$\effs+$\gamma_0^3$\effsdrei& \e0+$\gamma_0^2$\effszwei  &$\gamma_4$\effs&$\gamma_3$\eff\\
$\gamma_1$&$\gamma_4$\eff&\enulld+$\gamma_0^2$\effzwei&$\gamma_0^1$\effs+$\gamma_0^3$\effsdrei\\
$\gamma_4$\eff&$\gamma_3$\effs&$\gamma_0^1$\eff+$\gamma_0^3$\effdrei&\e0+$\gamma_0^2$\effzwei
\end{tabular}
\right )}
\label{eq:s}
\end{eqnarray}
and

\begin{eqnarray}
\nonumber {\displaystyle S({\bf k})=}\\ {\displaystyle \left (
\begin{tabular}{cccc}
1+$s_0^2$\effzwei& $s_0^1$\eff+$s_0^3$\effdrei &0&0\\
$s_0^1$\effs+$s_0^3$\effsdrei&1+$s_0^2$\effszwei&0&0\\
0&0&1+$s_0^2$\effzwei&$s_0^1$\effs+$s_0^3$\effsdrei\\
0&0&$s_0^1$\eff+$s_0^3$\effdrei & 1+$s_0^2$\effzwei
\end{tabular}
\right )}
.\label{eq:s}
\end{eqnarray}

\subsection*{Trilayer graphene}

The trilayer Hamiltonian and overlap matrices read as
\begin{eqnarray}
\nonumber {\displaystyle H({\bf k})=}\\ {\displaystyle \left (
\begin{tabular}{cccccc}
\enulld+$\gamma_0^2$\effzwei&\eff$\gamma_0^1$+$\gamma_0^3$\effdrei &$\gamma_1$&$\gamma_4$\effs&$\gamma_5$&0\\
$\gamma_0^1$\effs+$\gamma_0^3$\effsdrei&\e0+$\gamma_0^2$\effszwei&$\gamma_4$\effs&$\gamma_3$\eff&0&$\gamma_2$\\
$\gamma_1$&$\gamma_4$\eff&\enulld+$\gamma_0^2$\effzwei&$\gamma_0^1$\effs+$\gamma_0^3$\effsdrei&$\gamma_1$&$\gamma_4$\eff\\
$\gamma_4$\eff&$\gamma_3$\effs&$\gamma_0^1$\eff+$\gamma_0^3$\effdrei& \e0+$\gamma_0^2$\effszwei &$\gamma_4$\eff&$\gamma_3$\effs\\
$\gamma_5$&0&$\gamma_1$&$\gamma_4$\effs&\enulld+$\gamma_0^2$\effzwei&$\gamma_0^1$\eff+$\gamma_0^3$\effdrei\\
0&$\gamma_2$&$\gamma_4$\effs&$\gamma_3$\eff&$\gamma_0^1$\effs+$\gamma_0^3$\effsdrei&\e0+$\gamma_0^2$\effszwei
\end{tabular}
\right )}
\label{eq:s}
\end{eqnarray}
and
\begin{eqnarray}
\nonumber {\displaystyle S({\bf k})=}\\ {\displaystyle \left (
\begin{tabular}{cccccc}
1+$s_0^2$\effzwei&\eff$s_0^1$+$s_0^3$\effdrei &0& 0 & 0 &0\\
$s_0^1$\effs+$s_0^3$\effsdrei&1+$s_0^2$\effszwei& 0 & 0 &0& 0\\
0 & 0 & 1+$s_0^2$\effzwei & $s_0^1$\effs+$s_0^3$\effsdrei& 0 & 0\\
0 & 0 & $s_0^1$\eff+$s_0^3$\effdrei & 1+$s_0^2$\effzwei & 0 & 0 \\
0 & 0 & 0 & 0 &\enulld+$s_0^2$\effszwei & $s_0^1$\eff+$s_0^3$\effdrei\\
0& 0 & 0 & 0 & $s_0^1$\effs+$s_0^3$\effsdrei & 1+$s_0^2$\effszwei
\end{tabular}
\right ).}
\label{eq:s}
\end{eqnarray}

\end{widetext}

The sum of the phase factors for first--,second-- and third nearest neighbours are given by
$f_1({\bf k})$, $f_2({\bf k})$ and $f_3({\bf k})$, respectively. Note that $f_1({\bf k})$ and
$f_3({\bf k})$ couple $A$ and $B$ atoms and $f_2({\bf k})$ describes the $AA$ and $BB$ interactions.

\begin{equation}
f_1({\bf k})=\exp\left(i\frac{k_xa_0}{2}\right )+2\exp\left
(-\frac{ik_xa_0}{2}\right )\cos\left(\frac{\sqd k_ya_0}{2}\right).
\label{eq:f}
\end{equation}

\begin{equation}
f_2({\bf k})=\sum_{\ell=1}^6\exp(i{\bf k}{\bf r}^2_\ell) \label{eq:f}
\end{equation}

\begin{equation}
f_3({\bf k})=\sum_{\ell=1}^3\exp(i{\bf k}{\bf r}^3_\ell) \label{eq:f}
\end{equation}

Here $r_\ell^2$ and $r_\ell^3$ are the vectors that connect the $A_1$ atom (see Fig.~1) with the
second nearest $B$ atoms (three atoms) and the third nearest $A$ atoms (six atoms), respectively.

\end{document}